\documentclass[10pt,preprint]{emulateapj}
\shorttitle{White dwarfs in Close Binaries}
\shortauthors{Kawka et al.}

\begin{document}

\title{Orbital Parameters and Chemical Composition of Four
White Dwarfs in Post-Common Envelope Binaries}

\author{Adela Kawka\altaffilmark{1}, St\'ephane Vennes\altaffilmark{2}, 
Jean Dupuis\altaffilmark{3}, Pierre Chayer\altaffilmark{4}, Thierry Lanz\altaffilmark{5} }
\email{kawka@sunstel.asu.cas.cz, svennes@fit.edu, \\
jean.dupuis@space.gc.ca, chayer@pha.jhu.edu, tlanz@umd.edu}

\altaffiltext{1}{Astronomick\'y \'ustav AV {\v C}R,
Fri{\v c}ova 298, CZ-251 65 Ond{\v r}ejov, Czech Republic}
\altaffiltext{2}{Department of Physics and Space Sciences,
Florida Institute of Technology, Melbourne, FL 32901, USA}
\altaffiltext{3}{Canadian Space Agency, 6767 Route de l'Aeroport, Saint-Hubert,
QC J3Y8Y9, Canada}
\altaffiltext{4}{Department of Physics and Astronomy,
Johns Hopkins University, Baltimore, MD 21218}
\altaffiltext{5}{Department of Astronomy,
University of Maryland, College Park, MD 20742}
 
\begin{abstract}

We present $FUSE$ observations of the hot white dwarfs in the post-common 
envelope binaries Feige~24, EUVE~J0720$-$317, BPM~6502, and EUVE~J2013$+$400. 
The spectra show numerous photospheric absorption lines which trace the white 
dwarf orbital motion. We report the detection of \ion{C}{3}, \ion{O}{6}, 
\ion{P}{5}, and \ion{Si}{4} in the spectra of Feige~24, EUVE~J0720$-$317 and 
EUVE~J2013$+$400, and the detection of \ion{C}{3}, \ion{N}{2}, \ion{Si}{3}, 
\ion{Si}{4}, and \ion{Fe}{3} in the spectra of BPM~6502. Abundance measurements 
support the possibility that white dwarfs in post-common envelope binaries 
accrete material from the secondary star wind. The $FUSE$ observations of 
BPM~6502 and EUVE~J2013$+$400 cover a complete binary orbit. We used the 
{\it FUSE} spectra to measure the radial velocities traced by the white dwarf
in the four binaries, where the zero-point velocity were fixed using the
ISM velocities in the line of sight of the stellar systems. For BPM~6502 we
determined a white dwarf velocity semi-amplitude of 
$K_{\rm WD} = 18.6\pm0.5$km s$^{-1}$, and with the velocity semi-amplitude of
the red dwarf companion ($K_{\rm RD} = 75.2\pm3.1$km s$^{-1}$), we estimate
the mass ratio to be $q = 0.25\pm0.01$. Adopting a spectroscopic mass
determination for the white dwarf, we infer a low secondary mass of 
$M_{\rm RD} = 0.14\pm0.01\ M_\odot$. For EUVE~J2013$+$400 we determine a
white dwarf velocity semi-amplitude of $K_{\rm WD} = 36.7\pm0.7$km s$^{-1}$.
The $FUSE$ observations of EUVE~J0720$-$317 cover approximately 30\% of the 
binary period and combined with the $HST$ GHRS measurements 
(Vennes et al. 1999, ApJ 523, 386), we update the binary properties. 
{\it FUSE} observations of Feige~24 cover approximately 60\% of the orbit
and we combine this data set with $HST$ STIS (Vennes et al. 2000, ApJ, 544, 423)
data to update the binary properties.

\end{abstract} 

\keywords{stars: abundances --- binaries: spectroscopic
--- white dwarfs}

\section{Introduction}

Post common-envelope binaries consist of an evolved primary (white dwarf or
sdB) and a late-type main-sequence secondary in close orbit. These binaries
are thought to have evolved from wide binary systems, where the more massive
star evolved off the main sequence filling its Roche lobe and beginning mass
transfer onto its less massive companion. If the transfer is dynamically
unstable, a common-envelope (CE) is formed and friction between the stellar 
components and the CE decreases the orbital separation and induces the 
ejection of the CE. Depending on the separation of the components, some of these
systems further evolve to become cataclysmic variables. From the sample of
well studied post-CE binaries \citep{sch2003}, approximately half will evolve
into cataclysmic variables within a Hubble time.

The atmosphere of a white dwarf a in close binary system usually displays
enhanced traces of heavy elements. \citet{ven1999} found that the large
abundance of carbon and helium as well as time variable helium abundance
in some white dwarfs provide evidence for on-going accretion from the
red dwarf.

We present high-resolution far-ultraviolet (FUV) spectroscopic observations of
Feige~24 \citep[PG~0232$+$035:][]{ven1994b,ven2000}, WD~0718$-$316 
\citep[EUVE~J0720$-$317, 2RE~J0720$-$318:][]{ven1994a,bar1995b}, 
WD~1042$-$690 \citep[BPM~6502:][]{kaw2000} and
WD~2011$+$398 \citep[EUVE~J2013$+$400, 2RE~J2013$+$400:][]{tho1994,bar1995a} 
in \S 2. We present the analyses of the {\it FUSE} spectra in \S 3, 
determine new orbital and stellar parameters of the four binary 
systems in \S 4 and measure the heavy element abundance in the white dwarf 
atmospheres in \S 5. We summarize in \S 6.
 
\section{Observations}

\begin{figure*}
\plottwo{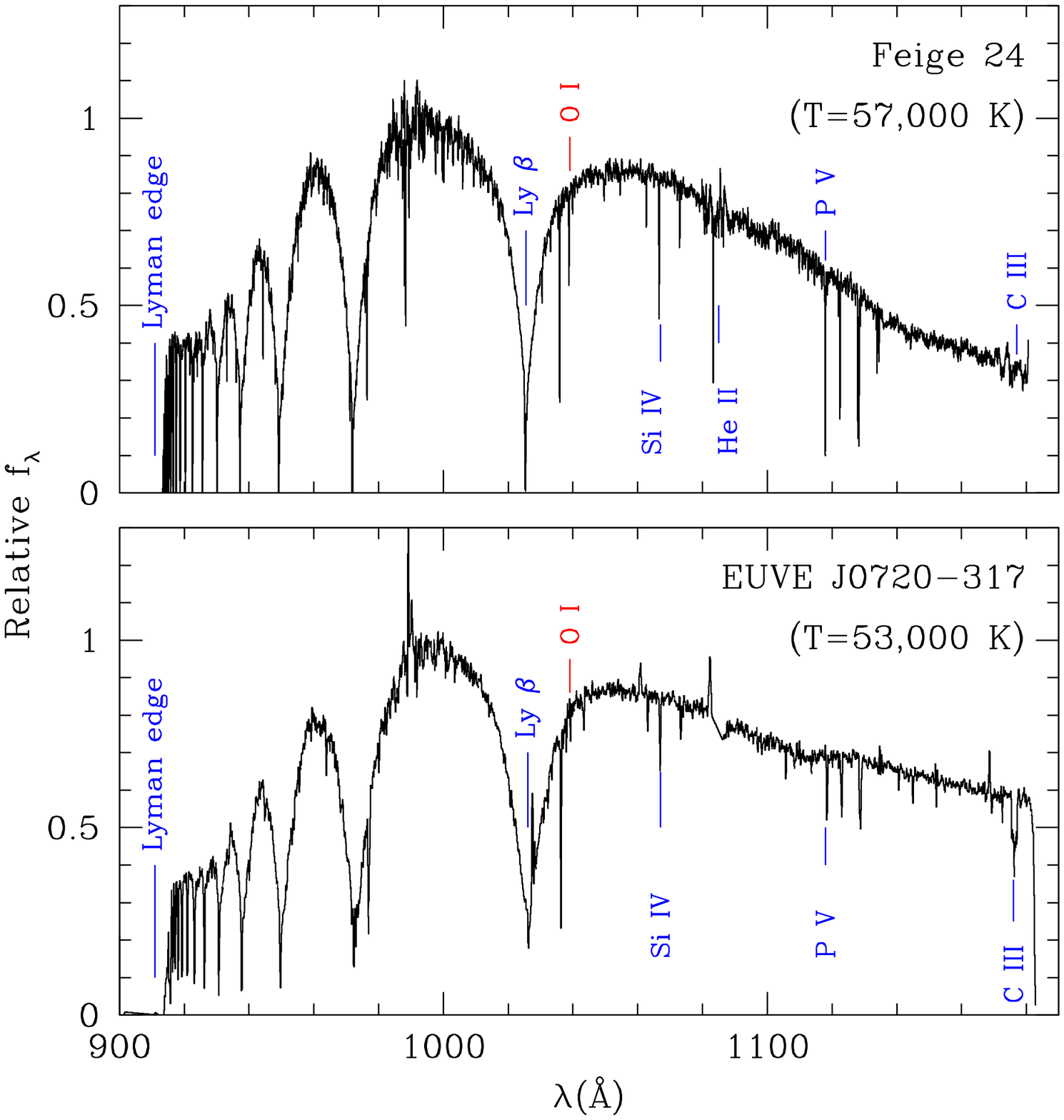}{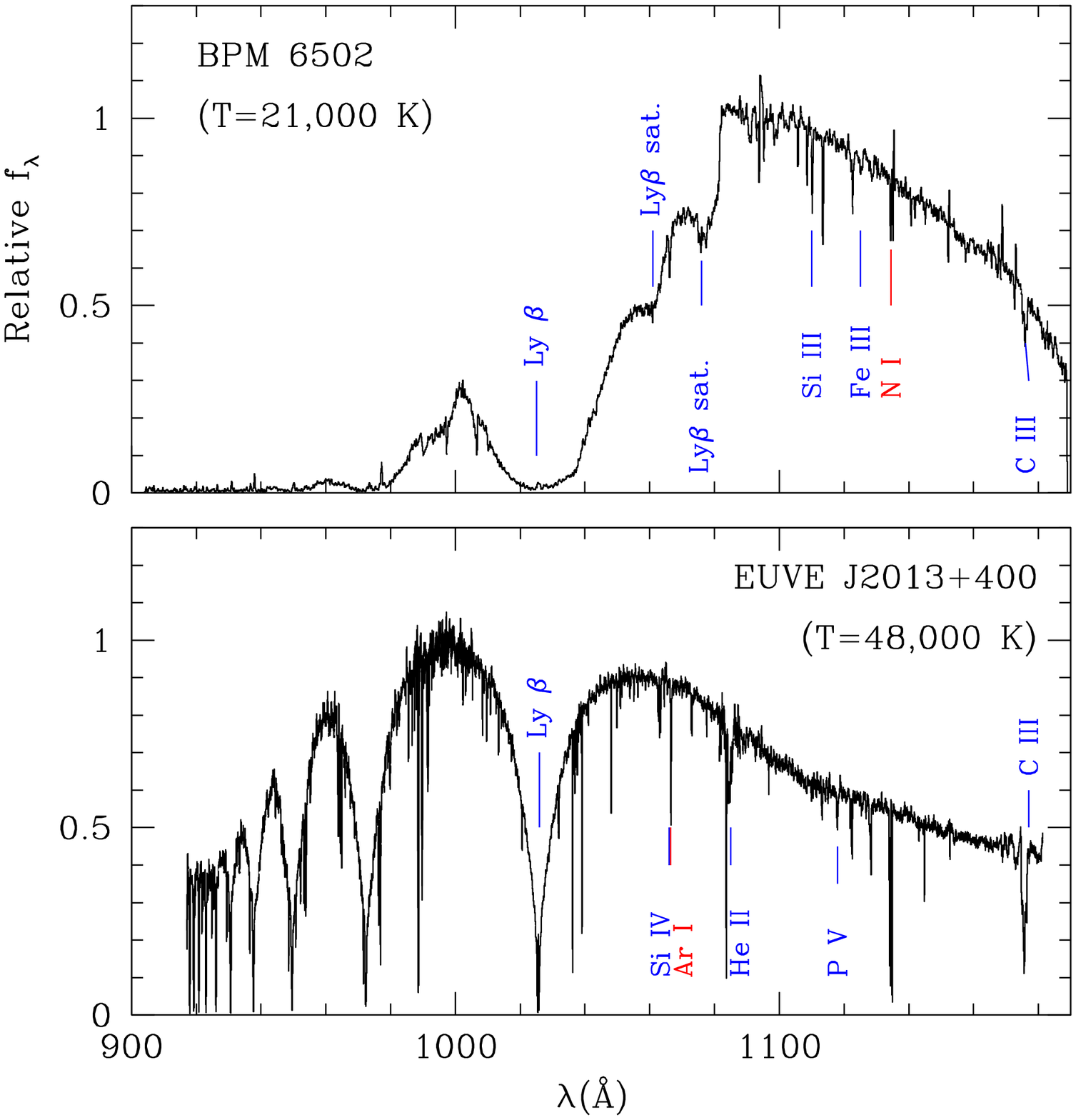}
\caption{{\it FUSE} spectra of Feige~24,  J0720$-$317, BPM~6502 and 
 J2013$+$400 showing the key heavy elements and ISM lines.
\label{fig_fuse_spec}}
\end{figure*}

We have obtained high-resolution FUV spectra of four close binary systems
with $FUSE$ (Table~\ref{tbl_fuse_log}). The
spectrograph covers the FUV spectral range from 905 to 1187 \AA\ with a
spectral resolution $R=20,000\pm2000$. The instruments are described in detail
by \citet{moo2000} and \citet{sah2000a,sah2000b}. The observations were made
using the LWRS (J0720$-$317, BPM~6502 and J2013$+$400) and MDRS (Feige~24, 
J2013$+$400) apertures and in time-tagged mode, except for Feige~24 which was 
observed in the HIST mode. The data for J0720$-$317 and BPM~6502 were processed 
with the CALFUSE pipeline v3.1 and the data for Feige~24 and J2013$+$400 were processed with
the CALFUSE pipeline v3.0. For the abundance analysis, we co-added the
individual exposures after aligning them on the photospheric lines using the
calculated radial velocities. Figure~\ref{fig_fuse_spec} shows the {\it FUSE} 
spectra of Feige~24, J0720$-$317, BPM~6502 and J2013$+$400 indicating key heavy 
elements. Table~\ref{tbl_feige24_lines} lists important photospheric lines in Feige~24,
J0720$-$317, and
J2013$+$400, while Table~\ref{tbl_bpm6502_lines} lists important lines
in BPM~6502. 

\begin{deluxetable*}{lccccc}
\tablewidth{0pt}
\tablecaption{$FUSE$ Observation Log \label{tbl_fuse_log}}
\tablehead{\colhead{Star} & \colhead{Date} & \colhead{Exposure time} & \colhead{Aperture} 
 & \colhead{Dataset} & \colhead{Observer}\\
                 &             &  (s)  &       &          &       }
\startdata
Feige 24         & 2003 Dec 07 &  2149 & MDRS  & P1040501 & Moos \\
                 & 2004 Jan 02 & 16117 & MDRS  & P1040503 & Moos \\
                 & 2004 Jan 06 & 11999 & MDRS  & P1040504 & Moos \\
  J0720$-$317 & 2001 Nov 13 & 17700 & LWRS  & B0510101 & Vennes \\
BPM 6502        &  2002 Jun 26 & 32100 & LWRS  & Z9104501 & Andersson \\
                 & 2006 Jan 21 & 10696 & LWRS  & U1074601 & Blair  \\
                 & 2006 Jan 22 &  7122 & LWRS  & U1074602 & Blair  \\
                 & 2007 Mar 02 & 12758 & LWRS  & U1074604 & Blair  \\
  J2013$+$400 & 2000 Oct 11 & 11200 & LWRS  & P2040401 & Moos \\ 
                 & 2002 Oct 29 &  6543 & LWRS  & M1053101 & Dupuis \\
                 & 2003 Oct 21 & 42607 & MDRS  & D0580101 & Vidal-Madjar \\
                 & 2003 Oct 23 & 12881 & LWRS  & M1053102 & Dupuis \\
\enddata
\end{deluxetable*}

\begin{deluxetable}{ll}
\tablewidth{0pt}
\tablecaption{Photospheric lines in Feige 24, J0720$-$317, and J2013$+$400.\label{tbl_feige24_lines}}
\tablehead{\colhead{Ion\ $\lambda$ (\AA)} & \colhead{Ion\ $\lambda$ (\AA)}}
\startdata
\ion{N}{4}   921.994  \ \ \ \ \ &  \ion{He}{2} 1084.94\tablenotemark{c} \ \ \ \ \\
\ion{N}{4}   922.519  & \ion{Si}{3} 1108.358 \\
\ion{N}{4}   923.676  & \ion{Si}{3} 1109.97\tablenotemark{d} \\
\ion{N}{4}   924.284  & \ion{Si}{3} 1113.23\tablenotemark{e}  \\
\ion{S}{6}   933.378  & \ion{P}{5} 1117.978  \\
\ion{S}{6}   944.523  &  \ion{P}{4}  1118.552$^a$ \\
\ion{P}{4}   950.657\tablenotemark{a} & \ion{S}{5}  1122.042 \\
\ion{N}{4}   955.334  & \ion{Si}{4} 1122.485  \\
\ion{C}{3}   977.020  & \ion{P}{5}  1128.008  \\
\ion{P}{4}  1030.515\tablenotemark{a} & \ion{Si}{4} 1128.34\tablenotemark{f} \\
\ion{O}{6}  1031.912  & \ion{S}{5}  1128.667  \\
\ion{P}{4}  1033.112\tablenotemark{a} & \ion{C}{3}  1174.993  \\
\ion{P}{4}  1035.516\tablenotemark{a} & \ion{C}{3}  1175.263   \\
\ion{O}{6}  1037.613  & \ion{C}{3}  1175.590  \\
\ion{S}{4}  1062.664  & \ion{C}{3}  1175.711  \\
\ion{Si}{4} 1066.63\tablenotemark{b}  & \ion{C}{3}  1175.987  \\
\ion{S}{4}  1072.973  & \ion{C}{3}  1176.370   \\
\ion{S}{4}  1073.518  & \nodata  
\enddata
\tablenotetext{a}{In Feige 24 only. \ion{P}{4} 950.657 may be blended with ISM \ion{O}{1} 950.885}
\tablenotetext{b}{Blend of $1066.614$, $1066.636$ and $1066.650$ \AA.}
\tablenotetext{c}{In J0720$-$317 and J2013$+$400 only.}
\tablenotetext{d}{Blend of $1109.940$ and $1109.970$ \AA.}
\tablenotetext{e}{Blend of $1113.174$, $1113.204$  and $1113.230$ \AA.}
\tablenotetext{f}{Blend of $1128.325$ and $1128.340$ \AA.}
\end{deluxetable}

\begin{deluxetable}{ll}
\tablewidth{0pt}
\tablecaption{Photospheric lines in BPM~6502.\label{tbl_bpm6502_lines}}
\tablehead{\colhead{Ion\ $\lambda$ (\AA)} & \colhead{Ion\ $\lambda$ (\AA)}}
\startdata
\ion{N}{2} 1083.990  & \ion{Fe}{3} 1128.050 \\
\ion{N}{2} 1084.576\tablenotemark{a} & \ion{Si}{4} 1128.34\tablenotemark{e} \\
\ion{N}{2} 1085.545\tablenotemark{b} & \ion{Fe}{3} 1128.724 \\
\ion{N}{2} 1085.701  &  \ion{Fe}{3} 1129.191 \\
\ion{Si}{3} 1108.358 &   \ion{C}{3} 1174.933 \\
\ion{Si}{3} 1109.97\tablenotemark{c} & \ion{C}{3} 1175.263 \\
\ion{Si}{3} 1113.23\tablenotemark{d} & \ion{C}{3} 1175.590 \\
\ion{Si}{4} 1122.485 & \ion{C}{3} 1175.711 \\
\ion{Fe}{3} 1122.526 & \ion{C}{3} 1175.987 \\
\ion{Fe}{3} 1124.881 & \ion{C}{3} 1176.370 \\
\ion{Fe}{3} 1126.729 & \nodata \\
\enddata
\tablenotetext{a}{Blend of 1084.562 and 1084.580 \AA.}
\tablenotetext{b}{Blend of 1085.529 and 1085.546 \AA.}
\tablenotetext{c}{Blend of $1109.940$ and $1109.970$ \AA.}
\tablenotetext{d}{Blend of $1113.174$, $1113.204$ and $1113.230$ \AA.}
\tablenotetext{e}{Blend of $1128.325$ and $1128.340$ \AA.}
\end{deluxetable}

\section{White Dwarf Atmospheric Parameters}

The {\it FUSE} spectrum of Feige~24 was analyzed by \citet{ven2005}. Using LTE models
they determined an effective temperature of $64700\pm3000$ K and 
$\log{g} = 7.58\pm0.25$. \citet{ven2005} also investigated the effect of
using LTE models on temperatures and surface gravities obtained from the
analysis of Lyman and Balmer lines of hot white dwarfs. They found
that for objects hotter 50\,000 K and assuming low metallicity, the LTE
determinations are overestimated by approximately 4000 K. Using the NLTE 
correction vector, the temperature would be revised to approximately 60\,000 K.
Which within uncertainties is in agreement with previous temperature
determinations, and in this work we will assume an effective temperature of
$T_{\rm eff} = 57\, 000\pm 2000$ K, which is based on estimates from previous
spectroscopic studies \citep[e.g.,][]{ven2001}.

The {\it FUSE} spectrum of BPM~6502 displays Lyman $\beta$ and $\gamma$ 
satellites. \citet{heb2003} compared the FUSE spectrum to a LTE model spectrum
at $T_{\rm eff} = 21\,380$ and $\log{g} = 7.86$ that included the 
quasi-molecular satellites of Lyman $\alpha$, $\beta$ and $\gamma$. The 
theoretical spectrum showed a reasonably good agreement with the observed 
{\it FUSE} spectrum of BPM~6502, however some discrepancies are observed.
\citet{heb2003} noted that their models do not include the variation of the
dipole moment during the collision, which may have a significant effect on the
strengths of the satellite profiles. In this work, we will adopt the 
effective temperature and surface gravity determined by \citet{kaw2007}. 
Even though these parameters were determined using optical spectra, only
H$\alpha$ is significantly contaminated, and as a result was excluded in their
analysis.

We used the {\it FUSE} spectra to obtain an effective temperature, surface
gravity and helium abundance of the white dwarfs in J0720$-$317 and J2013$+$400.
We computed a grid of LTE plane-parallel models. The grid of models extends
from $T_{\rm eff} = 30\,000$ to 70\,000 K (in steps of 4000 K), 
$\log{g} = 7.0$ to 9.5 (in steps of 0.25 dex) and 
$\log{(N_{\rm He}/N_{\rm H})} = -4.0$ to 0.0 (in steps of 0.5 dex). We fitted 6 
channels (SIC1B, SIC2A, LIF2B, LIF1A, SIC1A and SIC2B) simultaneously using a 
$\chi^2$ minimization technique to obtain an effective temperature, surface 
gravity and helium abundance. Regions which show interstellar absorption 
features were excluded from the fit. We obtained $T = 52\, 750\pm150$, 
$\log{g} = 7.73\pm0.03$ and $\log{(N_{\rm He}/N_{\rm H})} = -3.28\pm 0.08$ for 
J0720$-$317. For J2013$+$400 we obtained $T = 47\,800\pm200$, 
$\log{g} = 8.20\pm0.03$ and $\log{(N_{\rm He}/N_{\rm H})} = -2.90\pm 0.08$. 
The atmospheric parameters derived from {\it FUSE} spectra for J0720$-$317 are
consistent with parameters derived from Balmer line spectroscopic fits. 
The effective temperature and helium abundance for J2013$+$400 are consistent
with optical spectral analyses, however the surface gravity is significantly 
higher than the values determined from Balmer line spectral fits. This higher
surface gravity, which corresponds to a mass of $0.80\pm0.02\ M_\odot$ (using
the mass-radius relations of \citet{woo1995}) is also significantly higher than
the mass ($0.64\pm0.03\ M_\odot$) determined from the gravitational redshift of 
\citet{ven1999}. We will adopt the optical values in this work until this 
discrepancy is resolved. We discuss this problem further in \S 4.4.

Table~\ref{tbl_comp_par} summarizes the atmospheric properties of the white
dwarfs determined from {\it FUSE} spectra. The Table also compares these white dwarf parameters
to the same parameters based on optical spectroscopic studies.

\begin{deluxetable*}{lcccc}
\tablewidth{0pt}
\tabletypesize{\scriptsize}
\tablecaption{Properties of the white dwarf and its cool companion. \label{tbl_comp_par}}
\tablehead{
\colhead{Parameter\tablenotemark{a}} & \colhead{Feige 24} & \colhead{J0720$-$317} & \colhead{BPM 6502} & \colhead{J2013$+$400} }
\startdata
\multicolumn{5}{c}{White Dwarf}\\
\hline
$T_{\rm eff, opt}$ (K)   & $57\, 000\pm2000$\tablenotemark{b} & $52\, 400\pm 1800$\tablenotemark{c} & $19\, 960\pm400$\tablenotemark{d} & $48\, 000\pm 900$\tablenotemark{e} \\
$T_{\rm eff, FUSE}$ (K)  & \nodata\tablenotemark{f} & $52\, 750\pm 150$                   & \nodata\tablenotemark{g}          & $47\, 800\pm 200$                  \\
$\log{g}_{\rm opt}$ (c.g.s.)   & $7.66\pm0.08$                & $7.68\pm0.01$\tablenotemark{c}      & $7.86\pm0.09$\tablenotemark{d}    & $7.69\pm0.09$\tablenotemark{e}     \\
$\log{g}_{\rm FUSE}$ (c.g.s.)  & \nodata\tablenotemark{f} & $7.73\pm0.02$                     & \nodata\tablenotemark{g}          & $8.20\pm0.03$                      \\
$\log{(N_{\rm He}/N_{\rm H})}$ & \nodata                      & $-3.28\pm0.08$                      &  \nodata                          & $-2.90\pm0.08$                     \\
$M_{\rm opt}$ ($M_\odot$) & $0.58\pm0.05$\tablenotemark{h}    & $0.56\pm0.04$\tablenotemark{c}      & $0.55\pm0.05$                     & $0.56\pm0.03$\tablenotemark{e}     \\
$v_g$ (km s$^{-1}$)      & $20.1\pm 1.9$                      & $21.4\pm1.9$                        & $17.9\pm0.5$                      & $34.0\pm1.3$                       \\
$M_{\rm GR}$ ($M_\odot$) & $0.57\pm0.03$                      & $0.58\pm0.03$                       & $0.46\pm0.01$                     & $0.71\pm0.02$                      \\
\hline
\multicolumn{5}{c}{Red Dwarf}\\
\hline
M ($M_\odot$)    & $0.39\pm0.02$   & $0.43\pm0.03$ & $0.14\pm0.01$ & $0.23\pm0.01$ \\
R ($R_\odot$)    & $0.43\pm0.02$   & $0.47\pm0.03$ & $0.19\pm0.02$ & $0.29\pm0.01$ \\
$v_g$ (km s$^{-1}$) & $0.6\pm0.1$  & $0.6\pm0.1$   & $0.5\pm0.1$   & $0.5\pm0.1$   \\
Spec.Type.       & M2              &     M2        &     M5        &      M3       \\
\enddata
\tablenotetext{a}{All parameters are determined in this work unless indicated otherwise.}
\tablenotetext{b}{From \citet{ven2001}.}
\tablenotetext{c}{From \citet{ven1997b}.}
\tablenotetext{d}{From \citet{kaw2007}.}
\tablenotetext{e}{From \citet{ven1999}.}
\tablenotetext{f}{See text.}
\tablenotetext{g}{See \citet{heb2003} for a comparison between optical parameters and {\it FUSE} spectrophotometry (see text for details).}
\tablenotetext{h}{Determined using the measured parallax.}
\end{deluxetable*}

\section{Binary Parameters}

\begin{figure*}
\plottwo{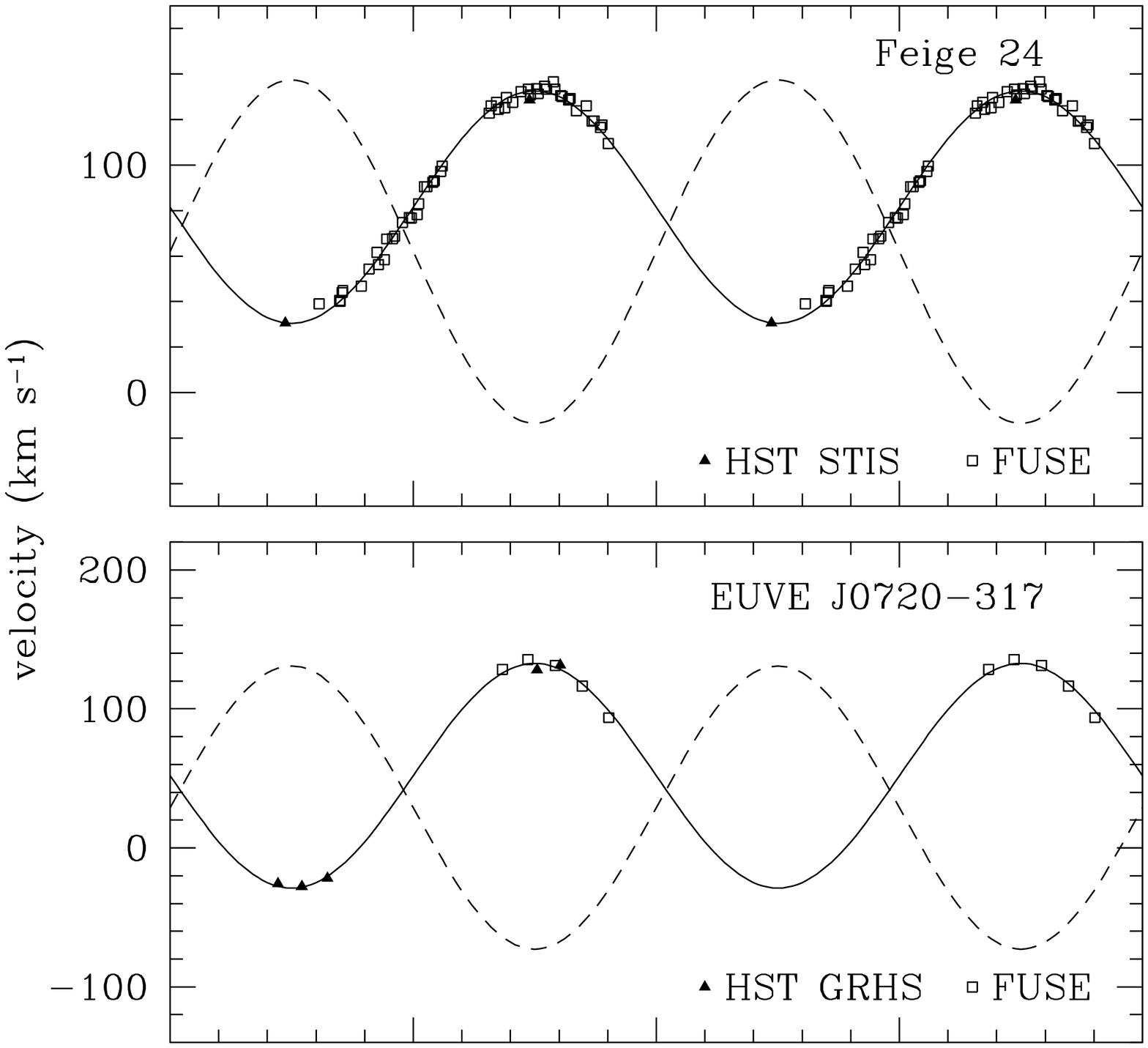}{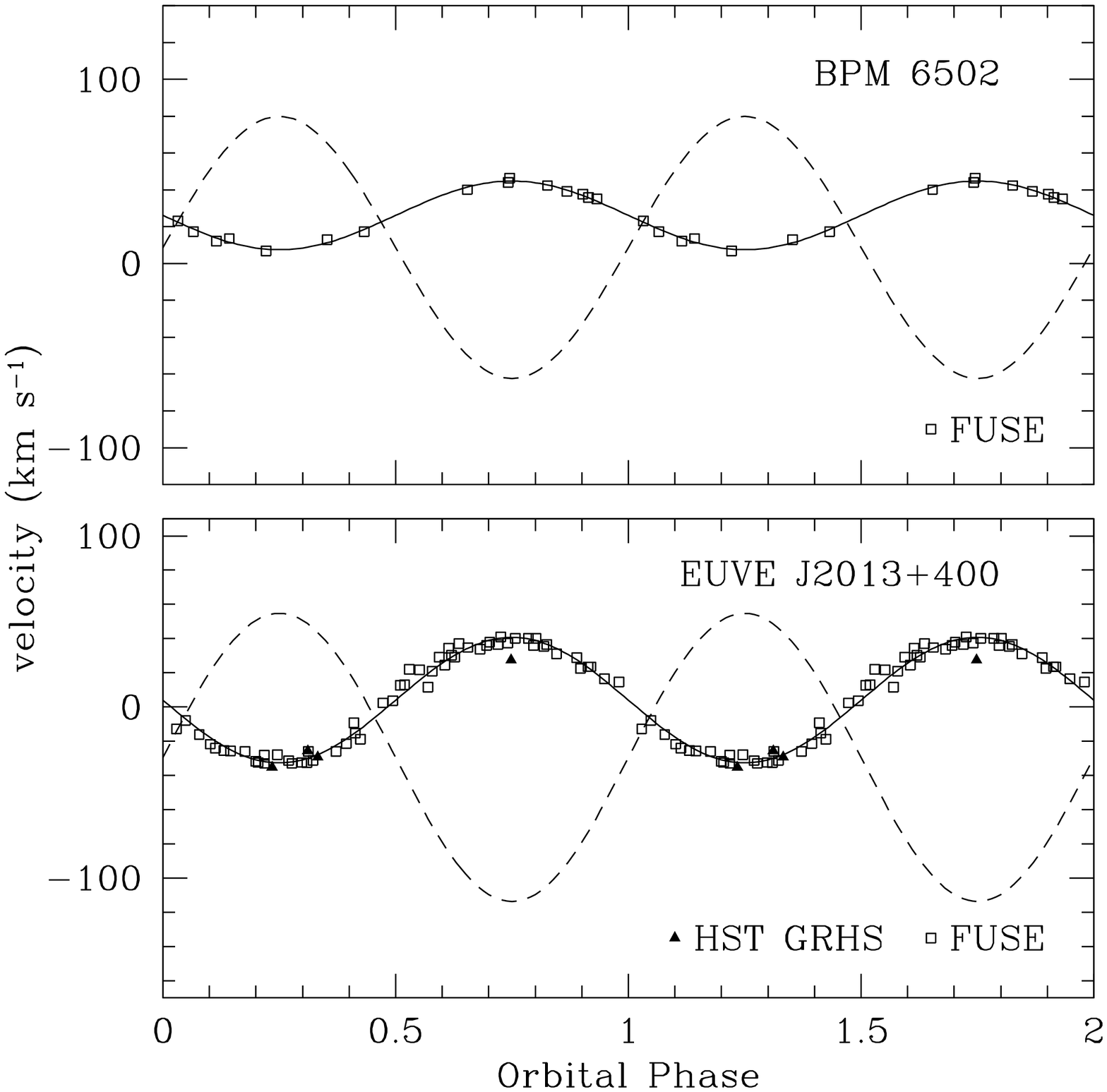}
\caption{Radial velocities of the white dwarf in Feige~24, J0720$-$317, 
BPM~6502 and J2013$+$400 folded on the orbital period as described in the 
text. \label{fig_wd_vel}}
\end{figure*}

We used the {\it FUSE} spectra to obtain radial velocities of the four
white dwarfs in the binary systems, Feige~24, J0720$-$317, BPM~6502 and 
J2013$+$400. These new radial velocity measurements are given in 
Table~\ref{tbl_fuse_vel}. We phased the white dwarf radial velocities using 
published binary ephemerides and determined the velocity
semi-amplitude $K_{\rm WD}$ and mean velocity $\gamma_{\rm WD}$. 
These new parameters are then combined with the results of previous optical studies
of the red dwarf companions which listed the red dwarf velocity semi-amplitudes 
($K_{\rm RD}$) and mean velocities ($\gamma_{\rm RD}$). From this set of 
measurements we calculated the binary mass ratios and the white dwarf gravitational 
redshifts ($v_{\rm g, WD}$) for all four systems. Figure~\ref{fig_wd_vel} shows the
radial velocity measurements of the four binaries folded on their orbital period.

\begin{deluxetable*}{ccc|ccc|ccc}
\tablewidth{0pt}
\tabletypesize{\scriptsize}
\tablecaption{{\it FUSE} Radial velocity measurements. \label{tbl_fuse_vel}}
\tablehead{\colhead{HJD} & \colhead{Phase} & \colhead{$v$} & \colhead{HJD}        & \colhead{Phase} & \colhead{$v$} & \colhead{HJD} & \colhead{Phase} & \colhead{$v$} \\
\colhead{(2450000+)} & \colhead{} & \colhead{(km s$^{-1}$)} & \colhead{(2450000+)} & \colhead{} & \colhead{(km s$^{-1}$)} & \colhead{(2450000+)} & \colhead{} & \colhead{(km s$^{-1}$)} }
\startdata
\multicolumn{9}{c}{Feige~24}\\
\hline
2980.556641 & 0.306 &  39.09 & 3007.908447 & 0.770 & 134.65 & 3010.681641 & 0.425 &  61.57 \\
2980.738525 & 0.349 &  40.54 & 3007.925049 & 0.774 & 133.33 & 3010.697021 & 0.429 &  56.18 \\
2980.743408 & 0.350 &  40.18 & 3007.987061 & 0.788 & 136.60 & 3010.749023 & 0.441 &  58.40 \\
2980.760498 & 0.354 &  43.99 & 3008.000000 & 0.792 & 133.43 & 3010.768555 & 0.446 &  67.56 \\
2980.765381 & 0.355 &  44.90 & 3008.049072 & 0.803 & 130.67 & 3010.820312 & 0.458 &  67.60 \\
3007.428223 & 0.656 & 122.77 & 3008.063721 & 0.807 & 130.21 & 3010.837158 & 0.462 &  68.66 \\
3007.443359 & 0.660 & 126.06 & 3008.118164 & 0.819 & 128.69 & 3010.905273 & 0.478 &  74.85 \\
3007.492432 & 0.672 & 127.64 & 3008.133545 & 0.823 & 129.36 & 3010.964844 & 0.492 &  76.92 \\
3007.509277 & 0.676 & 124.60 & 3008.188477 & 0.836 & 123.89 & 3010.980957 & 0.496 &  76.64 \\
3007.562744 & 0.688 & 125.24 & 3008.271729 & 0.856 & 126.05 & 3011.031006 & 0.508 &  78.30 \\
3007.578125 & 0.692 & 129.65 & 3008.325439 & 0.868 & 119.41 & 3011.044678 & 0.511 &  83.00 \\
3007.631592 & 0.704 & 127.55 & 3008.341064 & 0.872 & 119.50 & 3011.097412 & 0.523 &  90.51 \\
3007.706787 & 0.722 & 132.29 & 3008.394775 & 0.885 & 116.56 & 3011.118652 & 0.528 &  90.47 \\
3007.769775 & 0.737 & 133.51 & 3008.410400 & 0.889 & 117.65 & 3011.167725 & 0.540 &  92.46 \\
3007.786621 & 0.741 & 130.78 & 3008.462891 & 0.901 & 109.57 & 3011.183594 & 0.544 &  93.12 \\
3007.839355 & 0.754 & 133.54 & 3010.547607 & 0.393 &  46.78 & 3011.238281 & 0.557 &  97.22 \\
3007.855713 & 0.757 & 131.35 & 3010.613037 & 0.409 &  54.23 & 3011.251465 & 0.560 &  99.58 \\
\hline
\multicolumn{9}{c}{J0720$-$317}\\
\hline
2226.887805 & 0.682 & 125.4 & 2227.024896 & 0.791 & 127.5 & 2227.164429 & 0.902 & 93.7 \\
2226.954556 & 0.735 & 129.5 & 2227.095490 & 0.847 & 112.2 & \nodata     &\nodata&\nodata\\
\hline
\multicolumn{9}{c}{BPM~6502} \\
\hline
2452.329834 & 0.742 &  43.94 & 2452.636963 & 0.654 &  40.07 & 3758.507324 & 0.115 &  12.14 \\
2452.383789 & 0.902 &  37.55 & 2452.708984 & 0.868 &  39.26 & 4161.851074 & 0.745 &  46.36 \\
2452.438965 & 0.066 &  17.22 & 3757.399658 & 0.826 &  42.38 & 4161.914062 & 0.933 &  35.11 \\
2452.491455 & 0.222 &   6.87 & 3757.468994 & 0.032 &  23.22 & 4161.984863 & 0.143 &  13.60 \\
2452.562256 & 0.432 &  17.30 & 3758.439453 & 0.914 &  35.99 & 4162.055664 & 0.353 &  12.88 \\
\hline
\multicolumn{9}{c}{J2013$+$400} \\
\hline
1858.616821 & 0.261 & -31.31 & 2934.230957 & 0.833 &  31.12 & 2935.272705 & 0.309 & -31.28 \\
1858.688843 & 0.363 & -25.97 & 2934.303467 & 0.935 &  16.50 & 2935.344727 & 0.412 & -18.91 \\
1858.759399 & 0.463 &   2.42 & 2934.374268 & 0.036 &  -7.95 & 2935.411865 & 0.506 &  12.85 \\
1858.827881 & 0.560 &  11.69 & 2934.441895 & 0.132 & -25.66 & 2935.453857 & 0.566 &  21.00 \\
2576.795898 & 0.206 & -28.09 & 2934.483887 & 0.191 & -32.40 & 2935.482422 & 0.606 &  30.27 \\
2576.863037 & 0.301 & -25.85 & 2934.512695 & 0.232 & -27.90 & 2935.526123 & 0.668 &  33.93 \\
2576.932129 & 0.399 &  -9.26 & 2934.557617 & 0.296 & -32.68 & 2935.553223 & 0.707 &  36.46 \\
2577.002197 & 0.498 &  12.66 & 2934.631348 & 0.400 & -15.26 & 2935.599121 & 0.772 &  40.11 \\
2577.075684 & 0.603 &  34.30 & 2934.712891 & 0.516 &  22.12 & 2935.622559 & 0.805 &  35.33 \\
2936.028809 & 0.381 & -21.45 & 2934.767090 & 0.593 &  24.37 & 2935.672607 & 0.876 &  28.87 \\
2936.099121 & 0.481 &   3.57 & 2934.782471 & 0.614 &  29.15 & 2935.689697 & 0.900 &  23.53 \\
2936.169678 & 0.581 &  29.15 & 2934.834961 & 0.689 &  37.86 & 2935.736572 & 0.967 &  14.54 \\
2936.241455 & 0.682 &  35.89 & 2934.851807 & 0.712 &  40.96 & 2935.805664 & 0.065 & -16.10 \\
2936.312500 & 0.783 &  35.89 & 2934.904785 & 0.788 &  40.11 & 2935.822266 & 0.088 & -21.72 \\
2936.383301 & 0.884 &  22.68 & 2934.920898 & 0.811 &  36.46 & 2935.875000 & 0.163 & -25.94 \\
2934.023438 & 0.539 &  21.84 & 2934.987793 & 0.906 &  23.53 & 2935.891602 & 0.187 & -31.84 \\
2934.082764 & 0.623 &  37.02 & 2935.065674 & 0.016 & -12.73 & 2935.946045 & 0.264 & -32.96 \\
2934.096191 & 0.642 &  34.77 & 2935.125244 & 0.100 & -23.97 & 2935.960938 & 0.285 & -32.40 \\
2934.156982 & 0.728 &  37.30 & 2935.137695 & 0.118 & -25.38 & \nodata     &\nodata&\nodata \\
2934.167969 & 0.744 &  40.11 & 2935.199219 & 0.205 & -32.96 & \nodata     &\nodata&\nodata \\
\enddata
\end{deluxetable*}

Two small corrections were applied to the red dwarf data.  First,
a determination of the systemic velocity $\gamma_{\rm sys}$ requires that we subtract an estimate of the gravitational
redshift of the red dwarf ($v_{\rm g, RD}$) from its mean velocity. Secondly, 
a correction is applied to the red dwarf velocity semi-amplitude
to take into account the non-uniform distribution of the H$\alpha$ emission
over the red dwarf surface. Both corrections require initial estimates of
the red dwarf mass and radius. These estimates could be obtained by adopting
a mass and radius corresponding to published companion spectral type, but we favor
the following approach.

The strength of the H$\alpha$ emission line has been observed to vary in all four
systems \citep{ven1994b,kaw2002,tho1994} and is maximum near superior conjunction. 
These cyclical variations are explained by the changing viewing angle of the 
irradiated red dwarf hemisphere over the binary period \citep{tho1978}. In these 
systems, the H$\alpha$ emission originates from the irradiated hemisphere and 
traces an orbit that is lower than the true center of mass orbit. 
The irradiating flux originates from the hot photosphere of the white dwarf 
(e.g., Feige 24), or, alternatively, from a hypothetical hot accretion region on the surface of
the white dwarf as proposed by \citet{max1998} in the case of GD~448. 

We correct for the irradiation effect using the formalism described by 
\citet{ven1999}. \citet{wad1988} and \citet{oro1999} discuss a calculation assuming 
an irradiated hemisphere that employs the same formalism. The correction term uses 
a preliminary estimate of the red dwarf radius obtained using a spectroscopic 
determination of the white dwarf mass and the uncorrected mass ratio. The red dwarf 
radius is then calculated using the mass radius relations of \citet{cai1990}. The 
procedure may be iterated at will. However, the initial correction to the mass 
ratio being of the order of a few percent (5 to 8\%), further iterations are futile.

Final estimates of the red dwarf mass and radius are calculated from the corrected mass ratio and
the mass of the white dwarf and from the mass-radius relations of
\citet{cai1990}. Next, the gravitational redshift of the red dwarf is calculated using
$v_g = 0.63608 M(M_\odot)/R(R_\odot)$ km s$^{-1}$ and is typically of the order of
$0.5$ km s$^{-1}$. Finally, the systemic velocity is calculated using
$\gamma_{\rm sys} = \gamma_{\rm RD} - v_{\rm g, RD}$. 
Recent measurements
of late-type main sequence stars using interferometry and eclipsing binaries
show a relatively large scatter in the mass-radius relations for these stars
\citep{ber2006,lop2007}. \citet{reb2007} show that a similar scatter is 
observed in the late-type main sequence companions to white dwarf stars.
However, the effect of 
this scatter on the gravitational redshift of the secondary is less than 
0.1 km s$^{-1}$ and does not affect our results. 

\begin{figure}
\plotone{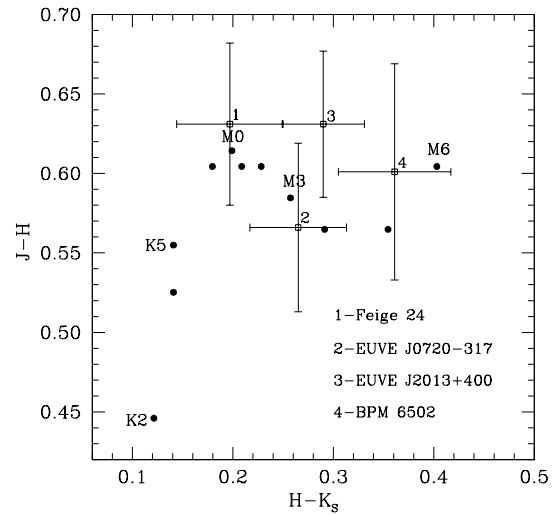}
\caption{2MASS $J-H$ versus $H-K_s$ color diagram showing the colors of
the secondary stars (corrected for the white dwarf contribution) in
Feige~24,  J0720$-$317,  J2013$+$400 and BPM~6502 ({\it open squares}), 
compared to main-sequence colors ({\it filled circles}). \label{fig_infra}}
\end{figure}

We estimate the spectral type of the secondary stars using the mass-type
relations of \citet{kir1994}, but we corroborate our spectral type determination
using 2MASS $JHK_s$ photometry. First, we calculate
the absolute $JHK$ magnitudes of the white dwarfs. These white dwarf 
magnitudes are then converted from CIT to 2MASS using \citet{cut2006}\footnote{Available at \\
http://www.ipac.caltech.edu/2mass/releases/allsky/doc/explsup.html}.
Next, we calculate the absolute magnitude of the system using the geometric parallax
(Feige 24) or an ultraviolet based photometric parallax of the white dwarf uncontaminated by the companion (J0720$-$317, 
BPM~6502 and J2013$+$400). Next, we subtract
the white dwarf contribution. Figure~\ref{fig_infra} shows the 2MASS $JHK_s$
of the secondary companions in the binary systems, Feige~24, J0720$-$317, 
BPM~6502 and J2013$+$400, compared to the colors for cool main-sequence stars
\citep{bes1988,bes1991} converted to 2MASS using \citet{cut2006}.

Finally, we determine the gravitational redshift of the white dwarf ($v_{\rm g}$) by 
subtracting the systemic velocity ($\gamma_{\rm sys}$) 
from the white dwarf mean velocity ($\gamma_{\rm WD}$).
Then, we convert the gravitational
redshift of the white dwarf into a mass estimate using mass-radius relations
of \citet{woo1995}.

With this general frame of work in mind, we now discuss each system separately.
The binary properties of these systems are given in Table~\ref{tbl_bin_par}, 
while the properties of the component stars are given in 
Table~\ref{tbl_comp_par}.

\begin{deluxetable*}{lcccccccc}
\tablewidth{0pt}
\tabletypesize{\scriptsize}
\tablecaption{Binary Parameters \label{tbl_bin_par}}
\tablehead{
\colhead{Parameter} & \colhead{Feige 24} & \colhead{J0720$-$317} & \colhead{BPM 6502} & \colhead{J2013$+$400}}
\startdata
$m_V$                           & 12.4                & 14.0                  & 13.0                    & 14.8                  \\
Period (days)                   & $4.23160\pm0.00002$\tablenotemark{a} & $1.262396\pm0.000008$\tablenotemark{b} & $0.3367849\pm0.0000006$ & $0.705517\pm0.000006$\tablenotemark{c} \\
$K_{\rm RD}$ (km s$^{-1}$)      & $75.5\pm2.1$\tablenotemark{a,d} & $98.2\pm1.2$\tablenotemark{b} & $71.1\pm0.2$ & $84.2\pm0.9$\tablenotemark{c} \\
$K_{\rm RD,corr}$ (km s$^{-1}$) & \nodata             & $105.9\pm3.4$  & $75.2\pm3.1$          & $89.1\pm2.6$          \\
$\gamma_{\rm RD}$ (km s$^{-1}$) & $62.0\pm1.4$\tablenotemark{a} & $31.1\pm0.7$\tablenotemark{b} & $8.7\pm0.1$ & $-29.5\pm0.7$\tablenotemark{c} \\
$\gamma_{\rm sys}$ (km s$^{-1}$) & $61.4\pm1.5$ & $30.5\pm0.8$ & $8.2\pm0.2$ & $-30.0\pm0.8$ \\
$K_{\rm WD}$ (km s$^{-1}$)      & $51.0\pm0.5$        & $80.8\pm1.2$          & $18.6\pm0.5$            & $36.7\pm0.7$          \\
$\gamma_{\rm WD}$ (km s$^{-1}$) & $81.5\pm0.4$        &  $51.9\pm1.1$         & $26.1\pm0.3$            & $4.0\pm0.5$           \\
$q$                             & $0.68\pm0.02$       & $0.76\pm0.03$         & $0.25\pm0.01$           & $0.41\pm0.01$         \\
$i$ ($^\circ$)                  & $77.4\pm3.5$        & $72.2\pm1.8$          & $20.3\pm0.3$            & $34.7\pm0.5$          \\
$f(m_{\rm WD})$ ($M_\odot$)     & $0.189\pm0.016$     & $0.155\pm0.015$       & $0.015\pm0.002$         & $0.052\pm0.005$       \\
\enddata
\tablenotetext{a}{From \citet{ven1994b}.}
\tablenotetext{b}{From \citet{kaw2002}.}
\tablenotetext{c}{From \citet{ven1999}.}
\tablenotetext{d}{Based on red dwarf absorption velocities.}
\end{deluxetable*}

\subsection{Feige~24}

\citet{tho1978} were the first to obtain orbital parameters for Feige~24 by measuring
the red dwarf radial velocities using H$\alpha$ and \ion{He}{1} emission lines.
\citet{ven1994b} determined an orbital period of $4.23160\pm0.00002$ days and
using the absorption lines they measured a red dwarf velocity semi-amplitude of $75.5\pm2.1$ km s$^{-1}$ with a
red dwarf mean velocity of $\gamma_{\rm RD} = 62.0\pm1.4$ km s$^{-1}$. 
Therefore, the measurements represent the true orbit traced by the secondary center of
mass. 

\citet{ven2000} obtained {\it HST} STIS spectra at orbital quadratures
and obtained an estimate of the white dwarf semi-amplitude of 
$49.1\pm0.3$ km s$^{-1}$ and a white dwarf mean velocity of $79.6\pm2.3$ km s$^{-1}$.
We measured the radial velocities of the white dwarf using the \ion{Si}{4} 
$\lambda 1066.63$ \AA\ line in the {\it FUSE} spectra. We used the ISM line of 
\ion{O}{1} $\lambda 1039.230$ \AA\ to fix the zero-point of the wavelength
calibration. We fixed the velocity of the ISM to $+7.4$ km s$^{-1}$, 
which is the mean velocity of the two ISM components in the line of sight
of Feige~24 \citep{ven2000}. We phased the {\it FUSE} data to the orbital
ephemeris of \citet{ven1994b} to obtain a white dwarf semi-amplitude of
$51.2\pm0.6$ km s$^{-1}$ with a white dwarf mean velocity of $81.6\pm0.4$ km s$^{-1}$.
The {\it FUSE} and {\it HST} white dwarf mean velocities are in agreement within
uncertainties, however the white dwarf semi-amplitude determined using
{\it FUSE} is slightly larger than determined by {\it HST}. Note that the
{\it FUSE} spectra lacks coverage at one of the quadratures ($\Phi = 0.25$). 
Combining the {\it FUSE} and {\it HST} data, we obtain 
$K_{\rm WD}=51.0\pm0.5$ km s$^{-1}$ and $\gamma_{\rm WD}=81.5\pm0.4$ 
km s$^{-1}$.
Figure~\ref{fig_wd_vel} shows the radial velocity measurements folded on the
orbital period. 

A minimum white dwarf mass of $0.53\ M_\odot$ is obtained using Kepler's 
third law. \citet{ven2000} estimated a white dwarf mass of 
$0.55\pm0.02\ M_\odot$ using the parallax measurement of $\pi = 14.7\pm0.6$ mas
\citep{ben2000}, 
$V = 12.56\pm0.05$ for the white dwarf \citep{hol1986} and adopting 
$T_{\rm eff} = 56\, 000\pm1000$ K. 
Using an effective temperature of $57\, 000\pm 2000$ K we
obtain a mass of $0.58\pm0.05\ M_\odot$. Note that the increased error in the
temperature results in an increased error in the mass determination.

We determined the white dwarf gravitational redshift to be 
$v_g = 20.1\pm1.9$ km s$^{-1}$. The gravitational redshift of the red dwarf 
was assumed to be $0.6\pm0.1$ km s$^{-1}$ (see below). The gravitational 
redshift of the white dwarf corresponds to a mass of $0.57\pm0.03\ M_\odot$.

Since the two mass estimates agree, we adopt a mass of $0.57\pm0.03\ M_\odot$ 
for the white dwarf, which is
the weighted mean of the two mass determinations discussed above. The
corresponding surface gravity would be $\log{g} = 7.66\pm0.08$. The mass
ratio of the system is $q=K_{\rm WD}/K_{\rm RD} = M_{\rm RD}/M_{\rm WD} = 0.68$
which results in a mass of $0.39\pm0.02\ M_\odot$ for the secondary. And with 
a radius of $0.43\pm0.02\ R_\odot$, we obtain a gravitational redshift of
$0.6\pm0.1$ km s$^{-1}$. The mass function of 
$f(M_{\rm WD}) = 0.189\pm0.016$ implies the inclination of the system is
$i = 77.4\pm3.5^\circ$.

The derived mass ($0.39\ M_\odot$) translates to a spectral type of dM1.5-2
for the secondary star. Using the 2MASS photometry of the binary system
($J= 11.265\pm0.024$, $H=10.733\pm0.022$, $K_s=10.557\pm0.019$), we 
calculated the absolute magnitudes of the secondary, $M_J = 7.28$, $M_H = 6.65$,
and $M_K = 6.45$. In our calculations we used the parallax distance of 
$68.4\pm2.0$ pc \citep{ben2000}. Figure~\ref{fig_infra} shows that the 
spectral type of the secondary based on 2MASS colors ($J-H = 0.63$, 
$H-K_s = 0.20$) ranges from $\sim$ K6-M2.

The binary parameters and mass of the secondary ($0.39\ M_\odot$) suggests
that magnetic braking will be the main angular momentum loss for the system,
and using the equations from \citet{sch2003} and \citet{rit1986}, we find that
the secondary star will fill its Roche lobe and begin mass transfer with a
period of 0.164 days in $\sim 2.2\times 10^{11}$ years. Therefore, Feige~24 is
not representative of the progenitors of the current population of cataclysmic 
variables.

\subsection{J0720$-$317}

Identified as a post-CE binary by \citet{ven1994a}, \citet{kaw2002} measured an
orbital period for the binary of $1.262396\pm0.000008$ days and a velocity 
semi-amplitude of the red dwarf of $98.2\pm1.2$ km s$^{-1}$ with a red dwarf
mean velocity of $\gamma_{\rm RD} = 31.1\pm0.7$ km s$^{-1}$. 
\citet{ven1999} used {\it HST} spectra to 
trace the orbit of the white dwarf, and measured the white dwarf semi-amplitude 
to be $79.5\pm1.4$ km s$^{-1}$. We have measured the radial velocities of the 
white dwarf using the \ion{Si}{4} $\lambda 1066.63$ line in the {\it FUSE} 
spectra. We used the ISM line of \ion{O}{1} $\lambda 1039.230$ \AA\
to fix the zero-point of the wavelength calibration. The velocity of the local
interstellar cloud (LIC) in the direction of J0720$-$317 is 13.5 km s$^{-1}$
\citep{lal1995}, however the measured velocity of \ion{O}{1} $\lambda 1039.230$ 
\AA\ is 6.7 km s$^{-1}$. Therefore, all measured velocities were shifted by 
$+6.8$ km s$^{-1}$. Combining the {\it FUSE} 
velocity measurements with the {\it HST} velocities we updated the white dwarf 
semi-amplitude to $K_{\rm WD} = 80.8\pm1.2$ km s$^{-1}$ and 
$\gamma_{\rm WD} = 51.9\pm1.1$ km s$^{-1}$. Figure~\ref{fig_wd_vel} shows the 
radial velocities of the white dwarf folded on the orbital period.

We corrected the red dwarf semi-amplitude to
obtain $105.9\pm3.4$ km s$^{-1}$, from which we estimate a mass ratio of
$q = 0.76\pm0.03$. Using the white dwarf mass of $0.56\pm0.04\ M_\odot$ 
from \citet{ven1997} and our measured mass ratio, we estimate the mass for the 
red dwarf to be $M_{\rm RD} = 0.43\pm0.03 M_\odot$ with a radius of
$0.47\pm0.03\ R_\odot$. Adopting the corrected mass function
of $f(M_{\rm WD}) = 0.155\pm0.015\ M_\odot$, we find the inclination of the 
system to be $i = 72.2\pm 1.8^\circ$.

The derived mass translates to a spectral type of dM1-2 for the secondary star.
Using the 2MASS photometry of the binary system ($J = 13.253\pm0.025$, 
$H = 12.749\pm0.026$, $K_s = 12.502\pm0.027$), we calculated the absolute
magnitudes of the secondary to be $M_J = 6.95$, $M_H = 6.38$ and $M_K = 6.12$.
To estimate the distance toward J0720$-$317, we obtained short-wavelength
IUE spectra (SWP 54496,54497) and compared it to a synthetic spectrum with
the white dwarf parameters ($T_{\rm eff} = 52400$ K, $\log{g} = 7.68$). The 
distance ($d$) to J0720$-$317 would then be when the difference between the 
observed spectrum and the model spectrum ($F$) placed at given distance, 
i.e., $(R/d)^2 F$ is minimized. The radius of the white dwarf ($R$) is
calculated using the mass-radius relations of \citet{woo1995}.
Figure~\ref{fig_infra} shows that the spectral type of the
secondary based on 2MASS colors ($J-H = 0.56$, $H-K_S = 0.26$) is $\sim$ dM3.

We determined the white dwarf gravitational redshift to be
$v_g = 21.4\pm1.9$ km s$^{-1}$. A red dwarf gravitational redshift of 
$0.6\pm0.1$ km s$^{-1}$ was used. The gravitational
redshift of the white dwarf corresponds to a mass of $0.58\pm0.03\ M_\odot$
which is in agreement with the mass determined from a spectroscopic fit of 
the Balmer lines (Table~\ref{tbl_bin_par}).

The binary parameters and the relatively large secondary mass suggests that
magnetic braking will be the main angular momentum loss for the system, and
using the equations from \citet{sch2003} and \citet{rit1986}, we conclude that
the secondary star will fill its Roche lobe and begin mass transfer with a 
period of 0.175 days (4.21 hrs) in $\sim 3.2\times 10^{9}$ years. The contact
period of 3.31 hrs reported by \citet{ven1999} is shorter than the one we
determined in this work, because of a larger secondary mass and hence higher 
mass ratio than the values used in \citet{ven1999}. Due to the large mass ratio
dynamically unstable mass transfer will be initiated when the system comes 
into contact \citep{dek1992}.

\subsection{BPM~6502} 
 
\subsubsection{Orbital Period}

Identified as post-CE binary by \citet{kaw2000}, \citet{kaw2002} measured an
orbital period of $0.336784\pm0.000001$ days, however \citet{mor2005} report
a period of $0.337083\pm0.000001$ days that is significantly longer. The period
determined by \citet{kaw2002} corresponds to their third alias. We have 
reanalyzed these two data sets and we show that the correct orbital period of 
BPM~6502 is the one reported by \citet{kaw2002}.

\begin{figure}
\plotone{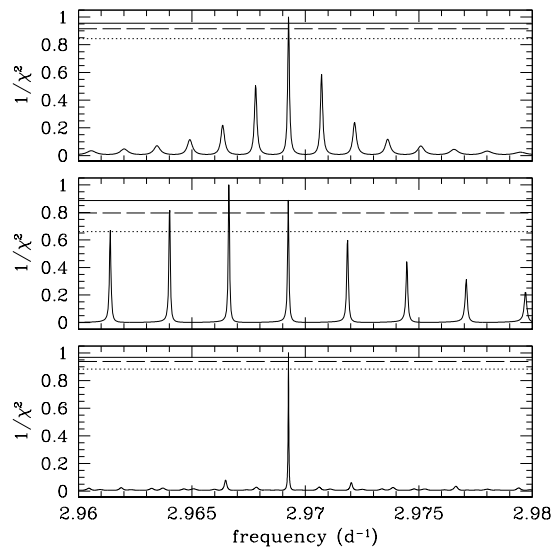}
\caption{{\it Top:} Periodogram of the radial velocities measured by 
\citet{kaw2002}. {\it Middle:} Periodogram of the radial velocities measured by
\citet{mor2005}. {\it Bottom:} Periodogram of the combined radial velocities. 
The lines shows the $1\sigma$ (full line), $2\sigma$ (dashed) and $3\sigma$ 
(dotted) confidence levels. \label{fig_per_bpm6502}}
\end{figure}

\begin{figure}
\plotone{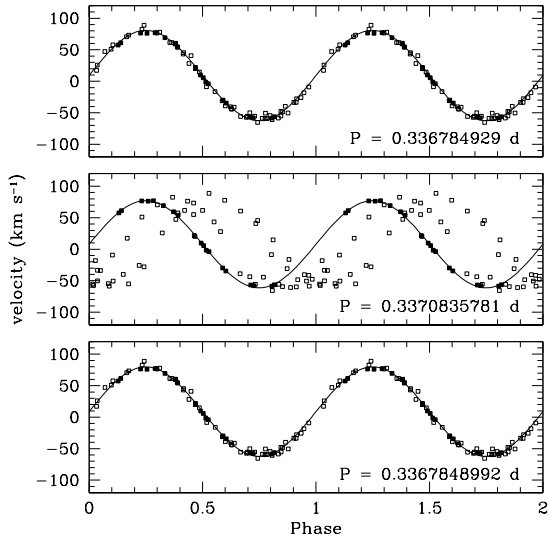}
\caption{Radial velocities of the red dwarf in BPM6502 traced by $H\alpha$ and
folded on the orbital period as determined by \citet{kaw2002} ({\it top}),
\citet{mor2005} ({\it middle}) and the period determined from combining the
radial velocity measurements from both studies ({\it bottom}). Velocities
measured by \citet{kaw2002} are indicated by open squares and velocities
measured by \citet{mor2005} are indicated by full squares. 
\label{fig_vel_bpm6502}}
\end{figure}

The first step of our analysis was to determine the period of the two sets
independently. The orbital period using measurements of \citet{kaw2002} is
$0.336784\pm 0.000001$ days, as reported. Aliases in the analysis are observed
but all of them are significantly below the $3\sigma$ confidence level. 
Figure~\ref{fig_per_bpm6502} (top panel) shows the periodogram of these 
measurements and Figure~\ref{fig_vel_bpm6502} (top panel) shows H$\alpha$ 
radial velocity measurements from both sets of data folded over the orbital 
period. Using the radial velocity measurements reported by \citet{mor2005}, we
reproduced the most probable orbital period of $0.337083\pm 0.000004$ days,
and the several aliases as reported in their analysis. We find
that 4 possible periods are within the $3\sigma$ confidence level, including 2
that fall within $1\sigma$. The second peak that falls within $1\sigma$ 
corresponds to a period of $0.336786$ days which  corresponds to the period
determined by \citet{kaw2002}. Therefore, using the data of \citet{mor2005} 
alone, the remaining 3 aliases cannot be excluded as possible orbital periods 
of the system. Figure~\ref{fig_per_bpm6502} (middle panel) shows the 
periodogram of these measurements, clearly showing that the other aliases 
cannot be excluded. The middle panel of Figure~\ref{fig_vel_bpm6502} shows 
H$\alpha$ radial velocity measurements from both sets of data folded over the 
best orbital period. Here the measurements taken during HJD 24510143 -
HJD 24510528 \citep{mor2005} follow the calculated orbital velocities,
however the measurements taken almost 3 years later HJD 2451607 - HJD 2451634
\citep{kaw2000} and a further 2 years later HJD 2452301 - HJD 2452317 
\citep{kaw2002} are out of phase.

Using both sets of data we found that the best orbital period is:
\begin{displaymath}
P = 0.3367849 \pm 0.0000006\ {\rm days}
\end{displaymath}
and the epoch of inferior conjunction
\begin{displaymath}
T_0 = 2450143.4195 \pm 0.0003
\end{displaymath}
where the new ephemeris is formally in agreement with \citet{kaw2002}.
Figure~\ref{fig_per_bpm6502} (bottom panel) shows the periodogram of these
measurements, where the aliases have almost disappeared. The bottom panel of
Figure~\ref{fig_vel_bpm6502} shows H$\alpha$ radial velocity measurements from
both sets of data folded over the best orbital period. The semi-amplitude
of the combined measurements is $K_{\rm WD} = 71.1\pm0.2$ km s$^{-1}$
with a red dwarf mean velocity, $\gamma_{\rm RD} = 8.7\pm0.1$ km s$^{-1}$. 

\subsubsection{White Dwarf Radial Velocity Measurements}

We measured the radial velocities of the white dwarf using the narrow 
\ion{Si}{3} lines $\lambda$1108.358, 1109.962, 1113.225 \AA\ in {\it FUSE} 
spectra. We used the ISM lines of 
\ion{N}{1} $\lambda 1134.165$, $1134.415$, $1134.980$ \AA\  to fix the 
zero-point of the wavelength calibration.
All sets of velocity measurements were corrected such that the ISM velocity is
equal to $v_{\rm ISM} = -7$ km s$^{-1}$ \citep{lal1995}. The semi-amplitude of 
the white dwarf radial velocity measurements is 
$K_{WD} = 18.6\pm0.5$ km s$^{-1}$ and 
$\gamma_{\rm WD} = 26.1\pm0.3$ km s$^{-1}$. Figure~\ref{fig_wd_vel} shows the 
radial velocities of the white dwarf folded over the period.

The H$\alpha$ emission in BPM~6502 is assumed to have originated from the
irradiated hemisphere of the red dwarf \citep{kaw2002}, and therefore we 
corrected the red dwarf semi-amplitude to $75.2\pm3.1$ km s$^{-1}$.
Using the corrected red dwarf semi-amplitude, we estimate a mass ratio of 
$q = 0.25\pm0.01$. The white dwarf mass of $0.55\pm0.05\ M_\odot$ was estimated 
from \citet{kaw2007} using the mass-radius relations of \citet{woo1995}.
From our measured mass ratio, we estimate the mass for the red dwarf to be
$M_{\rm RD} = 0.14\pm0.01 M_\odot$ and a radius of $0.19\pm0.02\ R_\odot$. 
Adopting the corrected mass function of
$f(M_{\rm WD}) = 0.015\pm0.002\ M_\odot$, we find the inclination of the system
to be $i = 20.3\pm 0.3^\circ$.

The derived mass ($0.14\ M_\odot$) translates to a spectral type of 
dM4.5($\pm0.5$). Using 2MASS photometry of BPM~6502 ($11.423\pm0.026$, 
$H = 10.896\pm0.027$, $K_s = 10.561\pm0.021$), we calculated the absolute
magnitudes of the secondary, $M_J = 8.88$, $M_H = 8.29$ and $M_K = 7.93$.
We determined the distance of BPM~6502 using IUE spectra (SWP 27351) in the 
same way as for J0720$-$317. Figure~\ref{fig_infra} shows that the spectral
type of the secondary based on 2MASS colors ($J-H = 0.60$, $H-K = 0.36$)
is $\sim$ dM5. \citet{tap2007} obtained K-band spectroscopy, where the spectral 
energy distribution suggests a spectral type ranging from M2.5 to M5, but the 
\ion{Na}{1} and \ion{Ca}{1} line strengths prefer a latter spectral type of M5. 
Taking into account all the available data we will adopt a spectral type of dM5.

We determined the white dwarf gravitational redshift to be 
$v_g = 17.9\pm 0.5$ km s$^{-1}$, which corresponds to a mass of 
$0.46\pm0.01\ M_\odot$. A red dwarf gravitational redshift of 
$v_g = 0.5\pm0.1$ km s$^{-1}$ was used. This mass is significantly different 
from the spectroscopic mass of $0.55\pm 0.05\  M_\odot$. Previous
gravitational redshift measurements of the white dwarfs using Balmer lines
resulted in much higher values, $42.2\pm9.0$ km s$^{-1}$ \citep{kaw2000}
and $36.16\pm2.72$km s$^{-1}$ \citep{mor2005}. It is likely that the Balmer
line cores are contaminated by the red dwarf even after the subtraction of
the emission line. In the case of using the {\it FUSE} spectra of the 
white dwarf to measure the radial velocity, the zero point depends on 
the adopted ISM velocity. We have assumed that the \ion{N}{1} lines originate
only from the LIC, however another cloud in the line of sight toward BPM~6502
could shift the velocity scale. Future high-dispersion ultraviolet observations of the
line of sight toward BPM~6502 may well be able to resolve velocity structures in ISM spectral
lines and provide a more reliable anchor for ultraviolet line velocities.

The binary parameters and the relatively small secondary mass suggests that
major contributor to angular momentum loss is from the release of gravitational 
radiation. Using the equations from \citet{rit1986}, the secondary
star will fill its Roche lobe and begin mass transfer with a period of
0.083 days in $\sim 3.0\times 10^{10}$ years. The time BPM~6502 comes into
contact is much longer than for J0720$-$317 because magnetic braking is not
invoked due to the low-mass of the secondary.

\subsection{J2013$+$400} 

Identified as post-CE binary by \citet{tho1994}, \citet{ven1999} measured an
orbital period for the binary of $0.705517\pm0.000006$ days and a semi-amplitude
of the red dwarf of $84.2\pm0.9$ km s$^{-1}$ with a mean velocity of
$\gamma_{\rm RD} = -29.5\pm0.7$ km s$^{-1}$. Using {\it HST} spectra they were able to trace the
orbit of the white dwarf, and measured the white dwarf semi-amplitude to be
$29.9\pm2.5$ km s$^{-1}$. We have measured the radial velocities of the white
dwarf using the \ion{Si}{4} $\lambda 1066.63$ line in the {\it FUSE} spectra. 
This line is near the ISM line of \ion{Ar}{1} $\lambda 1066.660$, and blends 
with the \ion{Si}{4} $\lambda 1066.63$ line for a short duration near orbital 
conjunction. We only included the
velocity measurements where the \ion{Si}{4} and \ion{Ar}{1} are clearly 
separated. \ion{Ar}{1} $\lambda 1066.660$ was used to fix the zero-point
of the wavelength calibration. Its proximity minimizes the distortion
in the wavelength scale due to thermal effects. All sets of velocity
measurements were corrected such that the ISM velocity is equal to 
$v_{\rm ISM} = -8$ km s$^{-1}$ \citep{lal1995}. Combining the {\it FUSE} 
velocity measurements with the {\it HST} velocities to obtain an updated white 
dwarf semi-amplitude of $K_{\rm WD} = 36.7\pm0.7$ km s$^{-1}$ and 
$\gamma_{\rm WD} = 4.0\pm0.5$ km s$^{-1}$. Figure~\ref{fig_wd_vel} shows the 
radial velocities of the white dwarf folded over the orbital period.

The H$\alpha$ emission is assumed to have originated from the irradiated 
hemisphere of the red dwarf, and hence does not trace the motion of the red
dwarf center of mass. Therefore, we corrected the red dwarf semi-amplitude to 
$89.1\pm 2.6$ km s$^{-1}$,
from which we estimate a mass ratio of $q = 0.41\pm0.01$. Using the white dwarf
mass of $0.56\pm0.03\ M_\odot$ from \citet{ven1999} and our measured mass 
ratio, we estimate the mass for the red dwarf to be
$M_{\rm RD} = 0.23\pm0.01 M_\odot$ and a radius of $0.29\pm0.01\ R_\odot$.
Adopting the corrected mass function of
$f(M_{\rm WD}) = 0.052\pm0.005\ M_\odot$, we find the inclination of the system
to be $i = 34.7\pm 0.5^\circ$.

The derived mass translates to a spectral type of dM3.5($\pm0.5$) for the 
secondary star. Using 2MASS photometry of J2013$+$400 ($J = 13.044\pm0.044$, 
$H = 12.520\pm0.024$, $K_s = 12.260\pm0.032$), we calculated the absolute
magnitudes of the secondary to be $M_J = 7.73$, $M_H = 7.09$ and $M_K = 6.80$.
We determined the distance of BPM~6502 using IUE spectra (SWP 27351) in the
same way as for J0720$-$317. Figure~\ref{fig_infra} shows that the spectral
type of the secondary based on 2MASS colors ($J-H = 0.64$, $H-K_S = 0.29$) is
$\sim$ dM4.

We determined the white dwarf gravitational redshift to be 
$v_g = 34.0\pm 1.3$ km s$^{-1}$, which corresponds to a mass of 
$0.71\pm0.02\ M_\odot$. A red dwarf gravitational redshift of
$v_g = 0.5\pm0.1$ km s$^{-1}$ was used. This is higher than 
the mass determined from a spectroscopic fit of the Balmer lines (see
Table~\ref{tbl_bin_par}) and than the mass of $0.64\pm0.03\ M_\odot$ determined
from the gravitational redshift by \citet{ven1999}. Again, as in the case of
BPM~6502, the possibility that the
adopted LIC velocity is blended with another ISM cloud causing a shift in the
velocity scale could explain this difference. Moreover, the mass determined from the 
{\it FUSE} spectral fit also indicates a higher mass ($0.80 M_\odot$). The
various mass estimates suggests that the white dwarf mass in J2013$+$400 is
is between $0.56\ M_\odot$ and $0.80\ M_\odot$.

The binary parameters and the relatively small secondary mass suggests that
major contributor to angular momentum loss is from the release of gravitational 
radiation. Using the equations from \citet{rit1986}, the secondary
star will fill its Roche lobe and begin mass transfer with a period of
0.118 days in $\sim 1.3\times 10^{11}$ years.

\section{White Dwarf Abundance Analysis}

\subsection{Model Atmospheres}

\begin{deluxetable*}{ccccccccccccc}
\tablewidth{0pt}
\tablecaption{Model atoms treated explicitly.\label{tbl_atoms}}
\tablehead{ & \multicolumn{5}{c}{BPM~6502} & &  \multicolumn{6}{c}{Feige 24, J0720$-$317, J2013$+$400} \\
\colhead{Element} & \colhead{I} & \colhead{II} & \colhead{III} &  \colhead{IV}  & \colhead{V} & & \colhead{I} & \colhead{II} & \colhead{III} & \colhead{IV} & \colhead{V} & \colhead{VI} \\
\cline{2-6} \cline{8-13} }
\startdata
 H\tablenotemark{a} & 9(1)    & \nodata & \nodata & \nodata & \nodata & & 9(1)    & \nodata & \nodata & \nodata & \nodata & \nodata \\
He\tablenotemark{a} & \nodata & \nodata & \nodata & \nodata & \nodata & & 24      & 20      & \nodata & \nodata & \nodata & \nodata \\
 C\tablenotemark{a} & \nodata & 22(5)   & 23(7)   & 25(4)   & \nodata & & \nodata & 22(5)   & 23(7)   &  25(4)  & \nodata & \nodata \\
 N\tablenotemark{a} & \nodata & 26(8)   & 32(7)   & 23(8)   & \nodata & & \nodata & \nodata & 32(7)   &  23(8)  & 16(6)   & \nodata \\
 O\tablenotemark{a} & \nodata & \nodata & \nodata & \nodata & \nodata & & \nodata & \nodata & \nodata & 39(8)   & 40(6)   & 20(5)   \\
Si\tablenotemark{a} & \nodata & 8\tablenotemark{b} & 30(6)  & 23(4) & \nodata & & \nodata & \nodata & 30(6) & 23(4) & \nodata & \nodata \\
 P\tablenotemark{b} & \nodata & \nodata & \nodata & \nodata & \nodata & & \nodata & \nodata & \nodata & 14      & 17(4)   & \nodata \\
 S\tablenotemark{b} & \nodata & \nodata & \nodata & \nodata & \nodata & & \nodata & \nodata & \nodata & 15      & 12      &  16  \\
Fe\tablenotemark{a} & \nodata & (36)\tablenotemark{c} & (50) & (43) & (42) & & \nodata & \nodata & \nodata & (43)  & (42) & (32) \\
\enddata
\tablenotetext{a}{{\it OSTAR2002}, \citet{lan2003}.}
\tablenotetext{b}{Earlier version}
\tablenotetext{c}{{\it BSTAR2006},  \citet{lan2007}.}
\end{deluxetable*}

We computed a series of NLTE model atmospheres using  TLUSTY version 200 and 
SYNSPEC version 48 \citep{hub1995} and adopted stellar parameters for BPM~6502,
J2013$+$400, J0720$-$317, and Feige~24: $T_{\rm eff}=20,000$K ($\log{g}=7.85$), 
and $T_{\rm eff}=48,000$ and $52,400$ K ($\log{g}=7.7$), and $57,000$K 
($\log{g}=7.5$), respectively. Using preliminary abundance estimates, we then 
bracketed the abundance inputs with models with abundance variations of $-0.7$, 
$0$ and $+0.7$ dex. Table~\ref{tbl_atoms} summarizes adopted model atoms, 
including some from \citet{lan2003,lan2007}. The table lists the number of 
levels included and, between parentheses, the number of super-levels. A 
super-level groups together several levels having close excitation energies and 
assumed to be in Boltzmann equilibrium relative to each other. For example, the 
ion \ion{C}{4} is treated using a total of 25 levels, four of them being 
super-levels. These four super-levels correspond to the principal quantum 
number of the external electron $n=7,8,9,10$, while detailed configurations are 
treated explicitly for $n=2$ (s, p) up to $n=6$ (s, p, d, f, g, h). The 
abundances are then measured using a $\chi^2$ minimization technique applied to 
observed spectral lines from Tables~\ref{tbl_feige24_lines} and 
\ref{tbl_bpm6502_lines} and the synthetic spectra.

In the case of J0720$-$317 and J2013$+$400 we also varied the iron abundance 
between $\log{\rm (Fe/H)}=-6$ and $-8$. The effect on the abundance of lighter 
elements did not exceed $\pm0.02$ dex. Therefore, we tabulate only abundances 
obtained with models at $\log{\rm (Fe/H)}=-6$. Finally, in the case of Feige~24 
we fixed the iron abundance to $\log{\rm (Fe/H)}=-5.5$ \citep{ven2000}.

We did not attempt vertical variations of the abundance of trace element. 
\citet{cha2003,cha2006} show evidence of a stratification of the oxygen 
abundance in several white dwarfs showing resonance lines of \ion{O}{6}. 
\citet{ven2001} also found a discrepancy between abundance measurements based 
on \ion{O}{4} and \ion{O}{5} in Feige~24 and G191~B2B which could be attributed 
to a stratification of oxygen in their atmospheres. 

\subsection{Analysis}

\begin{figure}
\plotone{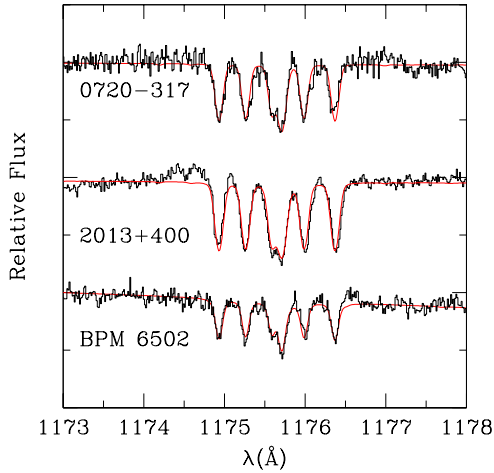}
\caption{{\it FUSE} spectra showing the \ion{C}{3} photospheric lines of
J0720$-$317, BPM~6502 and J2013$+$400 compared to the best fit 
models. \label{fig_spec_CIII}}
\end{figure}

\begin{deluxetable*}{ccccc}
\tablewidth{0pt}
\tablecaption{Photospheric Abundances for Feige 24, J0720$-$317, BPM~6502 and
J2013$+$400. \label{tbl_abun}}
\tablehead{\colhead{Element} & \multicolumn{4}{c}{$\log{\rm{(X/H)}}$} \\
\cline{2-5} \\
\colhead{} & \colhead{Feige~24} & \colhead{J0720$-$317} & \colhead{BPM~6502} &\colhead{J2013$+$400} }
\startdata
He & \nodata        & $-3.10\pm0.14$ & $\la -3$       & $-2.90\pm0.07$ \\
C  & $-6.90\pm0.06$ & $-5.75\pm0.08$ & $-5.68\pm0.07$ & $-5.50\pm0.05$ \\
N  & $-6.77\pm0.10$ & $-6.58\pm0.17$ & $-6.77\pm0.21$ & $-5.75\pm0.14$ \\
O  & \nodata \tablenotemark{a} & \nodata \tablenotemark{a} & \nodata & \nodata \tablenotemark{a} \\
Si & $-6.20\pm0.07$ & $-6.27\pm0.08$ & $-6.93\pm0.04$ & $-6.38\pm0.04$ \\
P  & $-7.11\pm0.05$ & $-7.50\pm0.10$ & \nodata        & $-8.20\pm0.05$ \\
S  & $-6.36\pm0.08$ & $-6.71\pm0.11$ & \nodata        & $-6.92\pm0.05$ \\
Fe & ($-5.5$)\tablenotemark{b} & \nodata        & $-7.32\pm0.13$ & \nodata \\
\enddata
\tablenotetext{a}{See text.}
\tablenotetext{b}{\citet{ven2000}.}
\end{deluxetable*}

Table~\ref{tbl_abun} lists the measured abundances in Feige~24, J0720$-$317, 
BPM~6502 and J2013$+$400. New {\it FUSE} abundance measurements in Feige~24 are 
consistent with the {\it HST} STIS measurements of \citet{ven2000}. 
Figure~\ref{fig_spec_CIII} shows \ion{C}{3} line profile fits. The carbon 
abundance measurements in J0720$-$317 and J2013$+$400 are in agreement with the 
measurements of \citet{ven1999} based on {\it HST} GHRS spectra of the 
\ion{C}{4} line profiles. We could not achieve satisfactory fits of the 
\ion{O}{6} line profiles using the present models. \citet{cha2003,cha2006} 
explored the problem in a sample of white dwarfs showing \ion{O}{6} resonance 
lines and concluded that a reservoir of oxygen concentrated at the top of the 
atmosphere is able to reproduce the line profiles with plausible oxygen 
abundance. We simply note that a similar phenomenon is possibly present in 
Feige~24, J0720$-$317, and J2013$+$400 and we exclude the case of oxygen from 
further consideration.

\begin{figure}
\plotone{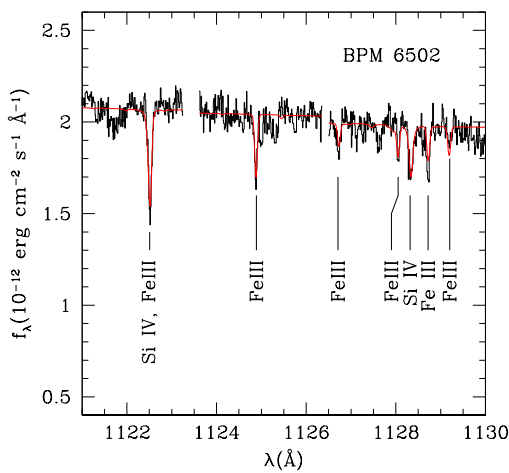}
\caption{{\it FUSE} spectra showing the \ion{Fe}{3} and \ion{Si}{4} (see 
Table~\ref{tbl_bpm6502_lines}) photospheric lines of
BPM~6502 compared to the best fit model. \label{fig_spec_FeIII}}
\end{figure}

Figure~\ref{fig_spec_FeIII} shows \ion{Si}{4} and \ion{Fe}{3} line profile 
fits in BPM~6502. At $T_{\rm eff}=20,000$K and $\log{\rm (Fe/H)}=-7.3$, 
BPM~6502 seems to follows a trend noted by \citet{ven2006} suggesting a 
decreasing abundance with decreasing temperatures. However, the abundance of 
iron in cooler white dwarfs may depend on their immediate environment. For 
example, the case of GD~362 at $T_{\rm eff}=9760$ K and $\log{\rm (Fe/H)}=-5.5$ 
\citep[or NLTT44986;][]{gia2004,kaw2006} may reveal the presence of 
circumstellar debris accreted onto the white dwarf. We could similarly argue 
that the presence of iron (and of lighter elements) in BPM~6502 is also a 
signature of accreted material but from the close late-type companion.

\begin{figure}
\plotone{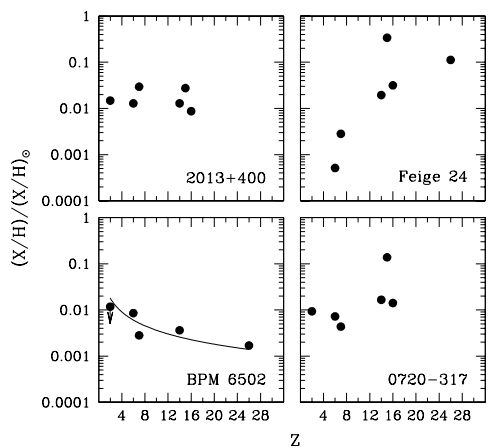}
\caption{Measured abundances in Feige 24,  J0720$-$317, BPM~6502 and
J2013$+$400 relative to solar values \citep{asp2005}. The iron abundance in 
Feige~24 is from \citet{ven2000}. \label{fig_abun_sol}}
\end{figure}

Figures~\ref{fig_abun_sol} summarizes our new abundance measurements in 
Feige~24, J0720$-$317, BPM~6502, and J2013$+$400. The abundance measurements 
($(X/H)/(X/H)_\odot$) are normalized to solar values \citep{asp2005} and 
presented as a function of the atomic number $Z$. Helium is dominant in 
J0720$-$317 and J2013$+$400, but the absence of \ion{He}{1} in 
medium-resolution spectra of BPM~6502 obtained by \citet{kaw2000} places a 
limit on the helium abundance of $\log{\rm He/H} \la -3$. Another interesting 
contrast emerges between Feige~24 and the three other binaries. The case of 
carbon is very instructive (see Figure~\ref{fig_spec_CIII}). The abundance of 
carbon is $\approx 1$\% solar in BPM~6502, J0720$-$317, and J2013$+$400, but it 
is $\approx 0.05$\% solar in Feige~24. Overall, Feige~24 presents an increased 
abundance toward larger atomic numbers. Although considerable scatter is 
present, one might conclude that, except for P in J0720$-$317 which appears 
somewhat more abundant, the abundances in BPM~6502, J0720$-$317, and 
J2013$+$400 are $\approx 0.2$-2\% solar. Especially in the case of J0720$-$317 
and J2013$+$400 this is in stark contrast with the abundance pattern noted in 
hot white dwarfs such as Feige~24 \citep[see][]{ven2001} which exhibit a yet 
unexplained trend showing increasing $(X/H)/(X/H)_\odot$ with $Z$. Although
selective radiation pressure, which is very effective in the case of elements
with a rich line spectrum (e.g., iron), plays an important role in white
dwarf atmospheres, detailed comparisons between theory and observations are not
satisfactory \citep{cha1995}.  In the cases 
of J0720$-$317 and J2013$+$400, the presence of helium (normally not present in 
hot DA white dwarfs) and of a concentration of heavy elements at $\approx 1$\% 
solar concentration suggest that these elements are accreted from the close 
late-type companion via wind/mass-loss. This wind/accretion scenario was 
explored in the case of the similar white dwarf plus late-type pair 1016$-$053 
\citep{ven1997b}. The white dwarf in Feige~24 being in a wider binary 
($P\approx 4.2$d versus $\la 1$ d) and considerably hotter bears more 
resemblance to hot, isolated DA white dwarfs such as G191-B2B \citep{ven2001}.

It is possible to estimate the accretion rate expected to bring about the
observed level of contamination. We adopt the accretion/diffusion model of 
\cite{mic1979} as a possible explanation for the abundance pattern observed in 
BPM~6502. The model may also be tentatively applied to the abundance of light 
elements in J0720$-$317 and J2013$+$400. This simple scenario implies that 
elements are accreted onto the white dwarf surface from the late-type wind; 
elements heavier than hydrogen, the main atmospheric constituent, diffuse 
downward at a certain rate. The equilibrium abundance at a certain point in 
the atmosphere is given by:
\begin{equation}
\frac{(X/H)}{(X/H)_\odot} = \frac{\dot{M}}{\dot{M}+4\pi R^2\rho v_d}
\end{equation}
where $(X/H)$ is the equilibrium abundance at that point, $(X/H)_\odot$ is the 
abundance in the red dwarf mass-loss (presumably solar), $\dot{M}$ is the 
accretion rate onto the white dwarf, $R$ is the white dwarf radius, $\rho$ is 
the density in the atmosphere at the abundance measurement point, and $v_d$ is 
the diffusion velocity at that same point. The white dwarf radius 
$R\approx 8.4\times10^8$cm is given by the stellar model, and for the present 
demonstration we will adopt the density $\rho$ at a Rosseland depth 
$\tau_R=2/3$ --- $\rho=7.3\times10^{-8}$ (for BPM~6502) and  
$\rho=1.4\times10^{-7}$ g cm$^{-3}$ (for J0720$-$317 and J2013$+$400) --- from
the atmospheric models discussed in \S 4.1. We then calculate the diffusion 
velocity $v_d$ at that location following the calculation of \citet{mic1979}. 
Table~\ref{tbl_diffuse} gives the diffusion velocity of key elements at 
$\tau_R=2/3$.

\begin{deluxetable}{ccc}
\tablewidth{0pt}
\tablecaption{Diffusion velocities at $\tau_R=2/3$.\label{tbl_diffuse}}
\tablehead{ \colhead{Element} & \multicolumn{2}{c}{$v_d$} \\
   & \multicolumn{2}{c}{(cm s$^{-1}$)} \\
\cline{2-3}\\
   & \colhead{BPM~6502} & \colhead{J0720$-$317, J2013$+$400} }
\startdata
\ion{He}{2} & \nodata & 0.003 \\
\ion{C}{3}  & 0.11    & 0.006 \\
\ion{Fe}{3} & 0.51    &\nodata \\
\enddata
\end{deluxetable}

First, we examine the case of BPM~6502 and estimate the accretion 
rate onto the white dwarf atmosphere using the carbon abundance 
$(X/H) = 0.01\times (X/H)_\odot$ and the diffusion velocity from 
Table~\ref{tbl_diffuse}.  A low rate of 
$\dot{M}= 1.1\times10^{-17} M_\odot {\rm yr}^{-1}$ is obtained. Next, assuming
that a constant iron to carbon ratio (solar) is preserved in the wind and 
assuming the same accretion rate, we determine the predicted equilibrium 
abundance using the correct diffusion velocity for iron. The predicted iron 
abundance is lower than the carbon abundance by the factor 
$v_d ({\rm C})/v_d({\rm Fe}) = 0.2$. Indeed, the abundance of iron in BPM~6502
is about a factor five lower than the carbon abundance. 
Figure~\ref{fig_abun_sol} shows that the expected trend 
$(X/H)/ (X/H)_\odot \propto A^{-1}$ (full line), where $A$ is the atomic 
weight, seems to apply to the case of BPM~6502. Renewed optical observations
of BPM~6502 aimed at placing a tighter constraint on the helium abundance 
would also help verify this accretion/diffusion model. The diffusion velocity
of singly ionized helium is similar to the velocity if doubly ionized carbon
which would imply similar equilibrium abundance. However, the diffusion velocity
of neutral helium is possibly a factor of 100 larger than for singly ionized
species which would considerably lower the equilibrium abundance of helium \citep{mic1978}.

A similar application of the accretion/diffusion model to the cases of 
J0720$-$317 and J2013$+$400 implies an accretion rate onto both white dwarfs of 
$\dot{M}= 1.8\times10^{-19} M_\odot {\rm yr}^{-1}$ (He) and 
$\dot{M}= 3.4\times10^{-19} M_\odot {\rm yr}^{-1}$ (C). The rates in the cases 
of J0720$-$317 and J2013$+$400 are markedly lower than in the case of BPM~6502
possibly because of their wider orbital separations or, hypothetically, their 
lower late-type mass-loss rates. \citet{ven1997b} determined a similar 
accretion rate onto the white dwarf 1016$-$053 of 
$\dot{M}= 8\times10^{-19} M_\odot {\rm yr}^{-1}$ based on their analysis of 
the helium  abundance in the EUV photosphere of that star. 

In summary, the observed abundance pattern in the atmosphere of BPM~6502 is 
consistent with an application of the accretion/diffusion model. The cases of 
J0720$-$317 and J2013$+$400 appear somewhat more complex, but their abundance 
patterns do bear some similarities with the pattern observed in BPM~6502.
Feige~24 being in a wider binary than the other binaries studied in our sample, 
the abundance pattern may be dominated by ordinary diffusion with possible 
effects due to radiative levitation and/or mass loss. In this context, it would 
be instructive to measure the iron abundance in the atmospheres of J0720$-$317 
and J2013$+$400. Such measurements can be obtained using far UV lines of 
\ion{Fe}{4} and \ion{Fe}{5} present at $\lambda \la 1300$ \AA.

\section{Summary}

We measured the white dwarf radial velocities in the close binary
systems Feige~24, J0720$-$317, BPM~6502 and J2013$+$400 using {\it FUSE}
spectra. Combined with previous optical studies of the binaries, we have
updated the binary properties. Of the four systems, only J0720$-$317
is likely to come into contact within a Hubble time and is therefore a
representative progenitor of the presently observed cataclysmic variables.
We have also reanalyzed the optical data of \citet{kaw2002} and \citet{mor2005}
and showed that the correct orbital period of BPM~6502 is 
$0.3367849\pm0.0000006$ days and that the period reported by \citet{mor2005}
should be discarded.

We measured the abundance of trace elements in the atmospheres of Feige~24, 
J0720$-$317, BPM~6502 and J2013$+$400 and found that the observed abundance 
pattern in BPM~6502 can be explained by steady accretion at a very low rate 
($\sim 1 \times 10^{-17}\ M_\odot$ yr$^{-1}$). In the cases of J0720$-$317 and
J2013$+$400 accretion from the secondary at much lower rates 
($\sim 10^{-19}\ M_\odot$ yr$^{-1}$) can explain the observed abundances,
however the higher abundances of P in these stars cannot be explained
by accretion from the secondary alone.

\acknowledgements

S. Vennes is grateful for the hospitality and support of the Astronomical 
Institute at Ondrejov Observatory. This work is supported by NASA grants 
NAG5-11717 and NAG5-6551. 
A.K. acknowledges support from the Grant Agency of the Czech Republic 
(GA \v{C}R) 205/05/P186 and the Centre for Theoretical Astrophysics. 
S.V. acknowledges support from the College of Science, Florida Institute of 
Technology. Based on observations made with the NASA-CNES-CSA Far Ultraviolet
Spectroscopic Explorer. FUSE is operated for NASA by Johns Hopkins University 
under NASA contract NAS5-32985.
This publication makes use of data products from the Two Micron All Sky Survey,
which is a joint project of the University of Massachusetts and the Infrared
Processing and Analysis Center/California Institute of Technology, funded by
the National Aeronautics and Space Administration and the National Science
Foundation.

\end{document}